# Less Is More: On the Importance of Data Quality for Unit Test Generation

JUNWEI ZHANG, The State Key Laboratory of Blockchain and Data Security, Zhejiang University, China

XING HU*, The State Key Laboratory of Blockchain and Data Security, Zhejiang University, China

SHAN GAO, Huawei, China

XIN XIA, Huawei, China

DAVID LO, Singapore Management University, Singapore

SHANPING LI, The State Key Laboratory of Blockchain and Data Security, Zhejiang University, China

Unit testing is crucial for software development and maintenance. Effective unit testing ensures and improves software quality, but writing unit tests is time-consuming and labor-intensive. Recent studies have proposed deep learning (DL) techniques or large language models (LLMs) to automate unit test generation. These models are usually trained or fine-tuned on large-scale datasets. Despite growing awareness of the importance of data quality, there has been limited research on the quality of datasets used for test generation. To bridge this gap, we systematically examine the impact of noise on the performance of learning-based test generation models. We first apply the open card sorting method to analyze the most popular and largest test generation dataset, Methods2Test, to categorize eight distinct types of noise. Further, we conduct detailed interviews with 17 domain experts to validate and assess the importance, reasonableness, and correctness of the noise taxonomy. Then, we propose CLEANTEST, an automated noise-cleaning framework designed to improve the quality of test generation datasets. CLEANTEST comprises three filters: a rule-based syntax filter, a rule-based relevance filter, and a model-based coverage filter. To evaluate its effectiveness, we apply CLEANTEST on two widely-used test generation datasets, i.e., Methods2Test and Atlas. Our findings indicate that 43.52% and 29.65% of datasets contain noise, highlighting its prevalence. Finally, we conduct comparative experiments using four LLMs (i.e., CodeBERT, AthenaTest, StarCoder, and CodeLlama7B) to assess the impact of noise on test generation performance. The results show that filtering noise positively influences the test generation ability of the models. Fine-tuning the four LLMs with the filtered Methods2Test dataset, on average, improves its performance by 67% in branch coverage, using the Defects4J benchmark. For the Atlas dataset, the four LLMs improve branch coverage by 39%. Additionally, filtering noise improves bug detection performance, resulting in a 21.42% increase in bugs detected by the generated tests.

---

*Corresponding Author

---

Authors' Contact Information: Junwei Zhang, The State Key Laboratory of Blockchain and Data Security, Zhejiang University, Hangzhou, China, jw.zhang@zju.edu.cn; Xing Hu, The State Key Laboratory of Blockchain and Data Security, Zhejiang University, Hangzhou, China, xinghu@zju.edu.cn; Shan Gao, Huawei, Hangzhou, China, gaoshan17@huawei.com; Xin Xia, Huawei, Hangzhou, China, xin.xia@acm.org; David Lo, Singapore Management University, Singapore, Singapore, davidlo@smu.edu.sg; Shanping Li, The State Key Laboratory of Blockchain and Data Security, Zhejiang University, 0000-0003-2615-9792, China, shan@zju.edu.cn.

# 1 Introduction

Unit testing is essential in software development to enhance the reliability and robustness of software applications [48]. However, manually writing high-quality test code is time-consuming and labor-intensive [5, 28, 33]. Various approaches are proposed to generate unit tests automatically, such as Randoop [42] and EvoSuite [19], to improve developers' productivity in writing tests.

Recently, the exploration of deep learning (DL) techniques, particularly large language models (LLMs) for generating unit tests, has demonstrated promising performance [1, 53, 61]. These methods are referred to as learning-based test generation approaches. They take the focal method (i.e., methods under test) as input to generate the unit tests. For instance, Alagarsamy et al. [1] pre-trained an LLM with focal methods and assertion statements, then fine-tuned it for test generation. Tufano et al. [62] proposed a similar method, pre-training LLMs with English and code corpora, followed by fine-tuning on test generation datasets. Despite the growing interest in learning-based test generation, little attention has been paid to the quality of datasets used to train these models. Noise in these datasets (e.g., irrelevant and erroneous) degrades the performance of test generation models. For example, noise may consist of code snippets with syntax errors or low-coverage test cases that fail to adequately test the focal methods. Such noise can result in faulty or inefficient test case generation, reducing the models' effectiveness. Previous work has shown that improving dataset quality can optimize DL models for various software tasks [55, 58]. For example, Sun et al. [58] proposed a data-cleaning framework for neural code search to improve the quality of code search datasets. Shi et al. [55] developed a taxonomy of data preprocessing noise for code summarization and built a rule-based cleaning tool to detect noise. However, few studies have investigated the quality of test generation datasets. Unlike other software tasks, unit test generation requires models to generate executable test cases that interact with focal methods. The noise patterns in test generation datasets differ from those in code search or summarization tasks because unit test generation must guarantee syntactic correctness, logical consistency, and code coverage. These requirements present unique challenges that have not been fully addressed in previous research. Therefore, it is crucial to conduct a deeper investigation into the quality of test generation datasets.

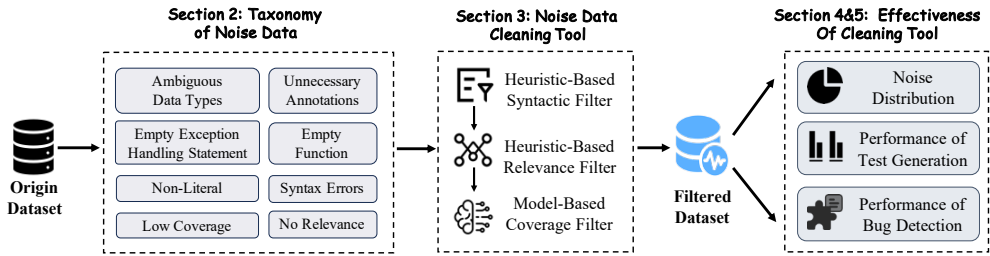

Fig. 1. An Overview of Our Research Methodology

In this work, we first investigate the noise within test generation datasets and the impact on learning-based test generation models. Specifically, we conduct a systematic study on the Methods2Test dataset [61], the latest and largest test generation dataset. Our research methodology is shown in Fig. 1. Rather than categorizing only noise data, we first label some datasets as noise based on specific characteristics. Then, we adopt the open card sorting method [52] to identify eight types of noise, including ambiguous data types, unnecessary annotations, empty exception-handling statements, missing implementation, syntax errors, non-English literal, irrelevant test cases, and low code coverage. Next, we conduct detailed interviews with 17 domain experts to validate the noise taxonomy in terms of reasonableness, correctness, and completeness. Based on this noise taxonomy, we propose a noise-cleaning framework, CleanTest. It systematically filters noise from the dataset using three key components: a rule-based syntactic filter, a rule-based relevance filter, and a model-based coverage filter. The rule-based syntactic filter applies a set of systematically designed heuristic rules to eliminate data with anomalous syntactic features, such as unnecessary annotations, empty exception-handling statements, and empty functions. The rule-based relevance filter selects data strongly correlated with the focal method and test cases. In addition to matching the function name called in the test code with the focal method's name [61], the number and type of parameters must also match. The model-based coverage filter refines the dataset by retaining data with higher branch coverage. The coverage filter relies on a pre-trained language model (PLM), fine-tuned with focal methods and test cases containing code coverage information from previous work [61]. We use the coverage filter to predict the code coverage of each instance and select those with higher branch coverage. We apply CleanTest to detect noise data and analyze the noise distribution in the Methods2Test and Atlas datasets. We observe that 43.52% of the Methods2Test dataset and 29.65% of the Atlas dataset consist of noise, highlighting its widespread presence. In Methods2Test, the unnecessary annotations noise constitutes the largest noise type of 41.64%, while the non-English literal noise accounts for the smallest of 0.16%. In Atlas, the syntax errors noise makes up the largest noise category at 28.74%. Other noise types, such as "ambiguous data type" (0.0051%), "unnecessary annotations" (0.0051%), "missing implementation" (0.0298%), "non-English literal" (0.0365%), and "No Relevance" (0%), have lower proportions.

To evaluate our noise-cleaning framework, we use the Defects4J benchmark [21] to compare the performance of four LLMs (i.e., CodeBERT [18], AthenaTest [61], StarCoder [38], and CodeLlama7B [51]) fine-tuned with datasets before and after filtering. Experimental results show that filtering noise data positively influences the test generation ability of models. In particular, the average performance of four LLMs in Methods2Test and Atlas datasets improves by 109.74% and 9.12% in CodeBLEU, 8.69% and 18.89% in syntactic correctness rate, 283.75% and 19.37% in compilation passing rate, 18.50% and 22.31% in line coverage, and 67.46% and 39.25% in branch coverage. Moreover, since less training data is used after filtering, we save at least 25% and 10% of training time with the same computational resources. We further evaluate the impact of noise data on the bug detection performance of the test generation model. The experimental results show that filtering noise improves the performance of bug detection, increasing the number of detected bugs by 21.42%. We also perform an ablation study to verify the effectiveness of each filter component and manually review the quality of the generated test cases. In summary, our main contributions are as follows:

- To the best of our knowledge, we are the first to systematically study the patterns and impact of noise in test generation datasets.
- We propose an automated noise-cleaning framework, named CleanTest, to detect and filter noise data in test generation datasets.
- We compare the performance of four LLMs fine-tuned on the original and filtered datasets using the Defects4J benchmark. The results demonstrate that filtering out noise yields significant performance improvements.
- For researchers and practitioners interested in our work, we release CleanTest and the filtered dataset for further research [44].

The remainder of this paper is organized as follows. Section 2 introduces the taxonomy of noise data in test generation datasets, categorizes eight distinct types of noise, and validates these categories through expert interviews. Section 3 presents our proposed noise-cleaning framework, CleanTest, which applies rule-based and model-based filters to remove noise data. Section 4 evaluates the impact of noise on the performance of four LLMs and demonstrates that filtering noise improves test generation effectiveness across multiple metrics. Section 5 analyzes the performance of CleanTest in bug detection and addresses threats to validity. Section 6 reviews related work on test generation and dataset quality. Finally, Section 7 summarizes the contributions of this work and proposes future research directions.

## 2 The Taxonomy of Noise Data

This section first presents the details of the Methods2Test and Atlas datasets. Second, we elaborate on the process of constructing the noise data taxonomy. Third, we define and provide examples of the eight identified noise types. Finally, we discuss the detailed interviews with 17 experts to validate the taxonomy.

### 2.1 Dataset Description

Our study conducts experiments on two widely used unit test generation datasets: Methods2Test [60] and Atlas [65]. These datasets primarily consist of pairs of focal methods and test cases, both represented at the method-level granularity. Table 1 summarizes the descriptive metadata and the related studies that have utilized each dataset in the literature.

Table 1. The Information of Methods2Test and Atlas Datasets

| Dataset | Year | Source | Language | #Pairs | #Projects | Trained-on Models |
|---|---|---|---|---|---|---|
| **Methods2Test** [60] | 2022 | GitHub | Java | 780,944 | 91,385 | [1, 15, 28, 60, 61] |
| **Atlas** [65] | 2020 | GitHub | Java | 188,154 | 9,275 | [1, 16, 40, 62, 65, 67, 75] |

Specifically, the Methods2Test dataset is the largest and most recent test generation dataset, containing metadata at various context levels. It includes 780,944 pairs from over 91,385 open-source Java projects, with updates ranging from 2017 to 2022. This dataset has been widely used in studies on unit test generation [1, 15, 28, 60, 61]. The Atlas [65] dataset contains 188,154 pairs from over 9,000 GitHub projects. Each test case in Atlas includes only a single assert statement. Besides, the dataset excludes all focal methods exceeding 1,000 tokens.

### 2.2 Taxonomy Construction

We adopt the open card sorting method [52] to categorize noise data for three main reasons: (1) It allows participants to create their categories rather than selecting from predefined options, providing the flexibility to adapt and refine categories as new insights emerge. (2) It supports iterative refinement, allowing assessors to revisit and adjust their classifications based on new information or feedback. Since the Methods2Test dataset is larger, more diverse, and unprocessed, we choose to conduct the noise type analysis on it. Then, we validate the distribution of different noise types in the Methods2Test and Atlas datasets in Section 4.3.

The open card sorting process consists of five rounds. In each round, we randomly sample 10,000 focal method and test case pairs from the Methods2Test dataset without replacement. In the first round, four participants follow three steps to label the sampled data. **Step ❶: Card Preparation.** A card is created for each focal method and test case pair, including the focal method, test case, and their respective file paths. **Step ❷: Card Classification.** Two researchers with at least three years of Java programming experience are invited. They independently analyze each card and determine whether the data represents noise or is "not noise". For data classified as noise, they assign specific noise categories and summarize keywords that describe the noise type. The researchers then discuss any differences in their classifications to reach a consensus on the labels. **Step ❽: Preference Voting for Answers.** To maintain diverse perspectives and minimize bias, two additional researchers with at least one year of software testing and three years of Java programming experience re-evaluate controversial cases. They classify the cards as either noise or not noise, and if classified as noise, assign the appropriate category. These assessors receive initial training to ensure alignment with the project goals and tasks.

The categories generated in the first round form a shared pool of categories. In subsequent rounds, two participants follow the same steps, either selecting an existing category from the shared pool or introducing a new one. The open card sorting process concludes when no new categories are added for two consecutive rounds. We conduct five rounds in total, labeling 50,000 pairs of focal methods and corresponding test cases. On average, the manual inspection of each instance requires approximately 30 seconds, amounting to a total of around 1,000 man-hours of dedicated effort. Finally, we identify and summarize eight potential noise types.

## 2.3 The Taxonomy of Noise Data

Based on the open card sorting process described above, we have identified eight types of noise. Below, we provide clear definitions and specific examples for each type.

↪ **Ambiguous Data Type.** This noise occurs when the focal method contains ambiguous data types. Unclear parameters and return values make it challenging for test generation models to understand the method's purpose and behavior, hindering the generation of effective test cases. Furthermore, generic methods provide significant flexibility, further complicating the testing of focal methods with ambiguous data types. For example, in Fig. 2, both the input parameter and return value in the "concat()" method are of type "Object".

```
public static Stream<Object> concat(Object... objects){
    ... }
```

Fig. 2. An Example of Ambiguous Data Types Noise

↪ **Unnecessary Annotations.** Annotations provide information about a program but do not contribute to its core functionality. Unnecessary annotation noise refers to annotations that play no significant role in the focal method's operation or functionality. For instance, in Fig. 3, the focal method "getPrefixes()" includes the "@ApiImplicitParams(...)" annotation, which describes implicit API documentation parameters. These annotations are primarily used by Swagger to generate API documentation and describe API parameters. Removing them does not affect the method's functionality or logic but only impacts the generation of API documentation and automatic description of parameters.

Unnecessary annotations consume valuable input space in LLMs, which have length restrictions. Long focal method tokens can lead to truncation and compromise the method's integrity. We calculate the token distribution of unnecessary annotations in the Methods2Test dataset, most annotations also fall into the [20, 30] token range, underscoring the need for filtering to enhance data relevance and model performance. These findings reinforce the necessity of filtering annotations to ensure dataset quality. Although some annotations may aid in understanding the code's logic and functionality, determining which are helpful and which are redundant is often subjective. To streamline the process, we remove all annotations, establishing a uniform and simplified noise removal procedure. This avoids the complexity of evaluating each annotation individually. Similarly, researchers in the Atlas dataset [65] also remove all annotations during dataset processing.

↪ **Empty Exception Handling Statement.** This noise type occurs when focal methods contain empty exception-handling statements (i.e., exceptions are caught but not handled), leading to bugs being ignored. For instance, in Fig. 4, the focal method "`from()`" includes an empty exception-handling statement. If a test generation model produces tests for such a noisy method, the test code may fail to detect and handle specific errors or exceptions, reducing test coverage and effectiveness.

Fig. 3. An Example of Unnecessary Annotation Noise

Fig. 4. An Example of Empty Exception Handling Statement Noise

**☞ Missing Implementation.** This noise type occurs when a method lacks implementation details, often because the function is defined elsewhere in the project, resulting in incomplete information in the dataset. When a focal method lacks logic or operations, it has no functionality to be tested. For example, in Fig. 5, the test case for the "`i18n()`" method requires a call to the "`msg()`" function, which is absent in the focal method. If the dataset contains such empty focal methods or test cases, the test generation model may replicate these patterns, resulting in numerous invalid test cases. Excluding this noise enables the model to generate more logically and functionally accurate code, enhancing the performance of fine-tuned LLMs.

Fig. 5. An Example of Missing Implementation Noise

**☞ Syntax Errors.** When the dataset contains syntax errors, such as missing semicolons or mismatched parentheses, the model may fail to parse the code structure of the focal method, resulting in incorrect test case generation. For instance, as shown in Fig. 6, a syntax error occurs in the focal method "`CompletableFuture()`" when it calls the "`setHref()`" function. Such errors can prevent fine-tuned LLMs from accurately parsing the code structure and generating correct test cases, thereby impairing their test generation capabilities.

Fig. 6. An Example of Syntax Error Noise

**☞ Non-English Literal.** This noise type occurs when the code contains strings in multiple languages, such as Japanese or Chinese, instead of English. Developers from different regions may write code in their native languages, mixing it with programming syntax. For example, in Fig. 7, the statement "`InvalidArgumentException`" includes Chinese characters. Although models can handle such names to some extent, non-standard naming conventions hinder their ability to recognize the structure and logic of the focal method, especially when the naming is closely related to the function's behavior. Furthermore, if the function processes strings with non-English characters, the model must correctly handle these characters to maintain functionality.

```java
public boolean resetPassword(Integer adminId, String password,String name) throws InvalidArgumentException{
    if(StringUtils.isBlank(name))
        throw new InvalidArgumentException("用户名不能空!");
... }
```

Fig. 7. An Example of Non-English Literal Noise

**⟳ No Relevance.** This noise type occurs when a test case calls a focal method different from the one intended to be tested. In Fig. 8, the focal method "`getWeight()`" and the test case "`testGetWeight()`" share similar names, but the focal method is not called in the test case. Instead, the test case is intended to test the "`getMatchingWeight()`" function in the "`SoundexMatcher`" class, which retrieves weights in the "`soundex`" class. This noise may result from ignoring method overloading when compiling the test generation dataset or focusing solely on name similarity without considering functional relevance. Removing such mismatches between focal methods and test cases ensures the model learns from data that is meaningful and contextually appropriate.

```java
Focal Method:
@Override public double getWeight(String str1, String str2){
    try{int diff = soundex.difference(str1,str2);
        return diff/MAX;}
    catch(Exception e){
        LOG.warn(e.getMessage(), e);
        return 0;}}
```

```java
Test case:
@Test public void testGetWeight(){
    SoundexMatcher soundexMatcher=new SoundexMatcher();
    String a = "John"; String b = "Jon";
    double matchingWeight=soundexMatcher.getMatchingWeight(a, b);
    assertEquals(1.0d, matchingWeight, EPSILON);
    a = "n"; b = "Hulme";
    matchingWeight = soundexMatcher.getMatchingWeight(a, b);
    assertTrue("not same "+a+" and "+b, 0.0d == matchingWeight);}
```

Fig. 8. An Example of the Noise with No Relevance Between Focal Method and Test Case

**⟳ Low Code Coverage.** This noise type arises in test cases with low code coverage. For instance, in Fig. 9, the "`convert()`" method processes the "`ExtractedDocumentMetadata`" object to convert its affiliations into a list of "`AffMatchAffiliation`" objects. The test "`convert_null_document()`" checks for a "`NullPointerException`" when the input is null. However, no test case checks the result when "`document.getAffiliations()`" is empty. Moreover, there are no test cases to verify the logic within the loop. Low-coverage tests may mislead the model into underestimating feature complexity or overestimating the robustness of the test suite.

## 2.4 Validation and Assessment

To validate and assess the noise taxonomy within the dataset, we conduct interviews with 17 software testing experts. Participants rate the reasonableness, correctness, and completeness of the taxonomy definitions on a Likert scale from 1 (low) to 5 (high). We use snowball sampling to invite experts [4], starting with an initial group who shared information about our study within their professional networks. We contact experts through emails, Twitter, and direct recruitment. The interviews are conducted from August 10, 2024 to August 17, 2024.

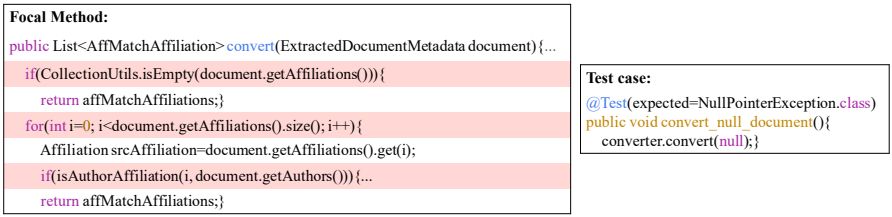

Fig. 9. An Example of Low Code Coverage Noise

The interviews are divided into three parts. **(1) Demographic information.** We ask some demographic questions, including geographical location, years of experience with LLMs and unit test generation, primary programming languages used, and the type of institution they are affiliated with. **(2) Statement scoring on the importance, reasonableness, and correctness of the noise taxonomy.** Experts rate these aspects on a 5-point Likert scale *(Strongly Disagree, Disagree, Neutral, Agree, and Strongly Agree* and an additional option *(I Do not Know)* [2]. The "*I Do not Know*" option includes statements that do not apply to experts' experience or are unclear to them. **(3) Rationale and suggestions.** In addition to scoring, interviewees would provide their rationale and suggestions for each statement and the completeness of the noise taxonomy.

We pilot the preliminary interview with two Ph.D. candidates to ensure its length and clarity. Based on their feedback, we make minor adjustments and produce the final version. The responses from the pilot interview are excluded from the final results. We analyze the interview results based on question types. For multiple-choice questions, we report the number of selections for each option. For open-ended questions, we use an inductive approach: two authors independently perform open card sorting and regularly discuss emerging themes until reaching a consensus.

The 17 experts are from two countries, i.e., China and Canada. Most experts (70.6%) have 1-2 years of experience using LLMs, 17.6% have 2-3 years. In terms of software testing experience, 10 experts (58.8%) have 1-2 years, 3 (17.6%) have 2-3 years, and 2 (11.8%) have 3-5 years of experience. Python is the most commonly used programming language among experts, with 88.2% indicating it as their primary language. Java follows with 70.6%, and C/C++ with 58.8%. Most experts (82.4%) are affiliated with the public sector, such as government or university institutions. 11.8% are in the private sector, and 5.9% are associated with the not-for-profit sector.

As shown in Table 2, we summarize 17 statements that capture the opinions of 17 experts regarding the assessment of the test generation dataset noise taxonomy. **(1) Importance (T1):** According to the results of S1, most experts rate the importance of dataset quality as 4 or 5 on a 5-point scale, with 58.8% giving a rating of 4 and 41.2% giving a rating of 5. This suggests that experts consider dataset quality crucial for improving test generation performance. **(2) Reasonableness (T2):** We summarize eight statements (S2-S9) regarding the reasonableness of various noise types in unit test generation. Overall, experts widely acknowledge and agree on the harmful effects of specific types of noise, particularly ambiguous data types, empty exception handling statements, missing implementations, syntax errors, no relevance, and low code coverage noises. Although 25% of experts remained neutral about unnecessary annotations and 30% about non-English literal noise, more than 50% believe that both types of data would introduce noise. **(3) Correctness (T3):** In Table 2, we summarize eight statements (S10-S17) to assess whether specific noise types are accurately defined. Mean ratings for these statements range from 3.76 to 4.53, suggesting that the definitions are accurate according to experts. A high level of agreement is observed for unnecessary annotations, empty exception handling statement, missing implementation, non-English literal, no-relevance, and low code coverage noises. **(4) Completeness:** Besides, we ask experts whether they believe there are other important types of noise not covered in the existing classifications, inviting them to elaborate if so. The responses indicate a unanimous consensus, as all experts answer "no" or "No, I do not", suggesting that the existing noise taxonomy is comprehensive.

Table 2. The Assessment of Test Generation Dataset Noise Taxonomy by 17 Domain Experts. (The Number of Respective Responses are Shown Using a Bar Plot)

| Statement | Likert Distributions | | |
|---|---|---|---|
| | Mean | Median | In Total |
| **T1. Importance of test generation dataset noise taxonomy** | | | |
| S1. Improving the quality of datasets benefits test generation models based on LLMs | 4.41 | 4.0 |  |
| **T2. Reasonableness of the definition of the test generation dataset noise type** | | | |
| S2. Including ambiguous data types would affect the quality of datasets | 4.00 | 4.0 |  |
| S3. Including unnecessary annotations would affect the quality of datasets | 4.00 | 4.0 |  |
| S4. Including empty exception handling statements would affect the quality of datasets | 3.82 | 4.0 |  |
| S5. Missing implementations in focal methods would affect the quality of datasets | 4.18 | 4.0 |  |
| S6. Including syntax errors would affect the quality of datasets | 4.12 | 4.0 |  |
| S7. Including non-English literal would affect the quality of datasets | 3.76 | 4.0 |  |
| S8. No relevance between focal methods and test cases would affect the quality of datasets | 4.35 | 4.0 |  |
| S9. Low code coverage between focal methods and test cases would affect the quality of datasets | 4.12 | 4.0 |  |
| **T3. Correctness of the definition of the test generation dataset noise type** | | | |
| S10. The content can be accurately described as the ambiguous data types noise | 3.76 | 4.0 |  |
| S11. The content can be accurately described as the unnecessary annotations noise | 4.12 | 4.0 |  |
| S12. The content can be accurately described as the empty exception handling statement noise | 4.29 | 4.0 |  |
| S13. The content can be accurately described as the missing implementation noise | 4.06 | 4.0 |  |
| S14. The content can be accurately described as the syntax errors noise | 3.82 | 4.0 |  |
| S15. The content can be accurately described as the non-English literal noise | 4.53 | 5.0 |  |
| S16. The content can be accurately described as the no relevance noise | 4.12 | 4.0 |  |
| S17. The content can be accurately described as the low code coverage noise | 4.12 | 4.0 |  |

## 3 The Noise Cleaning Framework

Building on the noise taxonomy introduced in Section 2, we propose CleanTest, an automated noise-cleaning framework. Insights from the taxonomy and expert interviews directly shape CleanTest's design, allowing it to target and eliminate the most harmful noise categories. CleanTest consists of three primary filters: a rule-based syntax filter, a rule-based relevance filter, and a model-based coverage filter. Each filter is specifically designed to address the noise types highlighted in the taxonomy, ensuring a more reliable dataset for training test generation models. In this section, we provide details of these three filters.

### 3.1 Syntax Filter

We follow previous work [55, 58] and use three criteria to construct heuristic rules: (1) Each rule should define a unique and specific type without overlap. (2) Rules should limit the exclusion of valid data within an acceptable range, with F1 scores above 90%. (3) No rule should be a sub-rule of another. We design a set of rules for each type of noise to detect it in the following three steps. **Step ❶:** Based on the manually annotated noise from Section 2, we carefully identify syntax features from 80% of the annotated data. **Step ❷:** We design a set of rules to detect these syntax features in the raw data. **Step ❸:** To avoid overfitting, we test the correctness of the rules on the remaining 20% of the annotated noise. We iteratively adjust the rules until the performance is acceptable.

We first parse focal methods and test cases into abstract syntax trees (ASTs) and use the depth-first search (DFS) algorithm [23] to traverse these ASTs. Then, we apply heuristic rules to filter out six types of noise. The rules for these six types of noise are as follows: **(1) For ambiguous data types noise**, we identify AST nodes of the "`method_declaration`" type and examine their "`type_parameters`" child nodes. If these nodes contain generic markers like $\langle E \rangle$, $\langle T \rangle$, $\langle K \rangle$, $\langle V \rangle$, $\langle N \rangle$, and $\langle ? \rangle$, we consider this data as noise and remove it. **(2) For noise data containing unnecessary annotations**, we identify "`method_declaration`" AST nodes and traverse their child nodes using DFS. If any child node is of the annotation type, we remove its text, considering it as noise. **(3) For noise data with empty exception handling statements**, we traverse the ASTs and check if the current node is of "`catch_clause`" or "`finally_clause`" type. If the child nodes are empty, we determine that the focal method is noise and remove the corresponding data from the original dataset. **(4) As for empty function noise**, we identify "`method_declaration`" nodes and traverse their child nodes. If these do not contain any further child nodes, we classify the method as empty function noise and remove it. **(5) For syntax error noise data**, we identify "`ERROR`" nodes in ASTs. If ASTs contain "`ERROR`" nodes, we consider this data as syntax error noise and remove it. **(6) For non-literal noise data**, we adopt regular expressions to identify this type of noise. The regular expressions are "$[\textbackslash uac00 - \textbackslash ud7ff]+$", "$[\textbackslash u4e00 - \textbackslash u9fa5]+$", and "$[\textbackslash u30a0 - \textbackslash u30ff \textbackslash u3040 - \textbackslash u309f]+$".

## 3.2 Relevance Filter

In the noise taxonomy, irrelevant test cases represent a significant noise type. This noise can severely degrade the quality of generated test cases, as the model might learn to associate irrelevant inputs and outputs. The relevance filter addresses this issue by retaining only test cases directly relevant to the focal method. It compares function signatures, parameter types, and parameter counts between the focal method and the test case. If a test case does not match the focal method based on these criteria, it is classified as irrelevant noise and removed. The steps of the relevance filter are detailed as follows:

**Step ❶: Parse focal methods and test cases into ASTs.** Both focal methods and test cases are parsed into their respective ASTs.

**Step ❷: Extract function signatures from the focal method.** We extract key information from the ASTs of focal methods, including the function name, number of parameters, and parameter types. Specifically, we adopt the DFS algorithm [23] to traverse the ASTs. When the AST node type is "`method_declaration`", we traverse its child nodes and extract the text of "`identifier`" and "`formal_parameters`" nodes as the function name and parameter types. The total number of parameters is counted to determine the number of parameters for the focal method.

**Step ❸: Extract called function information from the test case.** We first use the DFS algorithm to traverse the ASTs of test cases and extract all instances of function calls. For each function call, we use a process similar to Step ❷ to collect code details, including the name of the called function, the number of parameters passed, and their types.

**Step ❹: Validate information to determine relevance.** We verify the relevance of a matched function name by comparing the number of parameters and their types. Since a test case may contain multiple called functions, we only focus on those related to the focal method. We consider the focal method related to the test case only if at least one function call in the test case matches the focal method in name, number of parameters, and parameter types. If any of these aspects differ, we consider that there is no correlation between the focal method and the test case.

## 3.3 Low Coverage Filter

This filter aims to select the data with high branch coverage. It is an important metric that measures the percentage of tested branches out of the total available in the codebase [31]. Branch coverage inherently includes line coverage (if every branch is tested, every line in those branches must be executed) [66]. By focusing on branch coverage, test generation models are encouraged to generate test cases targeting all possible outcomes of conditional logic, thereby increasing their capability to catch bugs [31]. Since the Methods2Test dataset lacks code coverage information, we fine-tune CodeGPT [77] to predict the branch coverage of each focal method and test case pair in the Methods2Test dataset. The dataset provided by previous work [61] is used to fine-tune CodeGPT, including 1,269 focal methods and 25,680 test cases. Their code coverage information, obtained through coverage analysis tools [9], serves as labels. To avoid data leakage, we delete the data whose focal method names in this dataset are the same as the code names in the Methods2Test dataset. We randomly divide the code coverage dataset into 80%, 10%, and 10% for training, validating, and testing the coverage filter.

Specifically, given a focal method and a test case, we first concatenate these two code snippets. Then, we utilize the corresponding tokenizer of CodeGPT [77] to generate token sequences of the input. Following previous work [57], we use the hidden state of the last token as the code representation. Finally, a softmax layer is applied on top of the code representation to predict branch coverage. We train the coverage filter by minimizing the Mean Squared Error (MSE) loss. After predicting the branch coverage of the data and conducting a systematic inspection, we select those with branch coverage higher than 0.01 as the final filtered dataset. Since the test functions in the Methods2Test and Atlas datasets consist of individual code snippets rather than complete test suites, it is reasonable to select code snippets with higher coverage. By selecting snippets with higher coverage, it is possible to mitigate the limitations caused by the absence of complete test suites. This approach helps ensure the effectiveness of the testing process and enhances the robustness of the code.

To explore the impact of different PLMs on the performance of the coverage filter, we compare CodeGPT with two other PLMs: CodeBERT [18] and CodeT5 [64]. We use Mean Absolute Error (MAE) [63] and Mean Squared Error (MSE) [63] as evaluation metrics. Note that smaller values indicate better performance. As shown in Table 3, CodeGPT achieves the best performance (MAE: 7.98%, MSE: 1.05%) compared with other PLMs. Hence, we adopt CodeGPT as the coverage filter to predict branch coverage and select high-branch coverage data.

Table 3. The Performance Comparisons of Different PLMs in the Model-Based Coverage Filter

| Model Name | MAE | MSE |
|------------|--------|--------|
| **CodeBERT** | 36.52% | 14.38% |
| **CodeT5** | 9.76% | 1.43% |
| **CodeGPT** | **7.98%** | **1.05%** |

## 4 Experiment

In this section, we first present the research questions. Then, we describe our experimental setup, including the used datasets, the evaluated LLMs, evaluation metrics, and implementation details. Finally, we present the experiment results in this study.

### 4.1 Research Questions

We aim to answer the following research questions:
**RQ1:** What is the distribution of noise data in the Methods2Test and Atlas datasets?
**RQ2:** How effective is our noise-cleaning framework for unit test generation?
**RQ3:** What is the impact of each filter component on test generation?

### 4.2 Experimental Setup

*4.2.1 Datasets.* Our experiment involves two datasets: the training dataset and the validation dataset. The training dataset is used to fine-tune LLMs for test generation. The validation dataset, Defects4J [34], evaluates the performance of the fine-tuned LLMs. The statistics for both datasets are presented in Table 4.

Table 4. The Statistics of Three Datasets

| Type | Name | #Focal Methods | #Test Cases |
|---|---|---|---|
| **Training Dataset** | Methods2Test [60] | 586,468 | 780,944 |
| | Atlas [65] | 188,154 | 188,154 |
| **Validation Dataset** | Defects4J [34] | 757 | 1,440 |

✋ **Training Dataset.** As introduced in Section 2, we use the popular test generation datasets, Methods2Test [61] and Atlas [65], to fine-tune LLMs-based test generation models. For each of them, we shuffle these datasets and partition them into two parts, 80% for fine-tuning the LLMs and 20% for hyperparameter tuning. Since the projects in the validation dataset (i.e., Defects4J) are also public repositories on GitHub, it is challenging to determine if a project in Defects4J is not in the Methods2Test and Atlas datasets. This presents a major threat to the effectiveness evaluation [6, 59]. We mitigate this threat by removing focal methods in Methods2Test and Atlas datasets that share the same function names as those in the validation dataset.

✋ **Validation Dataset.** We utilize Defects4J [21, 34] to evaluate how well a test generation model performs in a real-world scenario [21, 34]. The Defects4J dataset contains 835 bugs from 17 Java open-source software systems. Following previous work [61], we extract pairs of focal methods and test cases. Finally, we collect 757 focal methods and 1,440 test cases. Three projects (i.e., JacksonCore [10], JacksonXml [13], and JxPath [35]) are excluded because the test case name differs from the focal method name after the "*Test*" word. The Defects4J benchmark is also processed using the CleanTest, ensuring the validation dataset undergoes the same noise-cleaning procedures for a more accurate evaluation of the four LLMs.

### 4.2.2 Test Generation Model.
This study aims to explore the impact of data quality on the performance of learning-based test generation models. Since LLMs are pre-trained with extensive code and natural language data, they have significant advantages in understanding and generating code. For our experiments, we select four LLMs (CodeBERT [18], AthenaTest [61], StarCoder [38], and CodeLlama7B [51]) commonly used in recent research [61, 74].

- **CodeBERT [18]:** CodeBERT is the most popular and widely-used code-specific LLMs. It pre-trains the transformer architecture with natural and programming languages. We use CodeBERT and a transformer-based decoder to generate tests.
- **AthenaTest [61]:** AthenaTest is a state-of-the-art model specifically designed for automated unit test generation. It is trained on the Methods2Test dataset [61], allowing it to learn complex relationships between focal methods and their corresponding unit tests. AthenaTest is fine-tuned using developer-written tests. Following previous work [76], we use CodeT5 [64] to replicate AthenaTest. CodeT5 is an encoder-decoder transformer model pre-trained on 5.2 million code functions and 8.3 million natural language sentences.
- **StarCoder [38]:** StarCoder is a state-of-the-art code LLM with 15.5B parameters, pre-trained on data from GitHub. We fine-tune it to generate test cases.
- **CodeLlama7B [51]:** CodeLlama7B is a more recent SOTA model based on the Llama2 architecture, specifically fine-tuned for code-related tasks. It is frequently used in recent studies focusing on code generation due to its large-scale pretraining and strong performance on various benchmarks.

### 4.2.3 Evaluation Metrics.
We adopt five widely used metrics in our experiments to evaluate the performance of four LLMs.

- **CodeBLEU [50]**: This metric compares the generated code with human-written code by considering both structural and semantic information, including the token match calculated by the BLEU [45], the syntax match calculated by AST structure similarities, and the data flow match calculated by data flow similarities. We set a weight of 0.25 for each aspect.
- **Syntactic Correctness Rate**: This metric verifies the syntactic correctness of the generated tests according to Java rules, using the tree-sitter parser [46, 69].
- **Compilation Passing Rate**: This metric represents the percentage of generated tests that execute without runtime errors, using JUnit as the test executor.
- **Line Coverage**: This metric quantifies the number of code lines executed by the test [27, 31]. We average the line coverage of all tests to obtain the final metric.
- **Branch Coverage:** This metric measures the proportion of branches in the focal method executed by the test [9, 31, 41]. We calculate the average rate of multiple tests as the final metric.

*4.2.4    Implementation Details.* Our experimental environment is equipped with four NVIDIA GeForce RTX A800 GPUs and 128 Intel(R) Xeon(R) CPUs, running on Ubuntu OS. When fine-tuning the four test generation models with the training dataset, we follow the implementations provided in their original papers and adopt the recommended hyper-parameter settings. Besides, we follow prior work [61] and adopt the same prompt as the input format to fine-tune LLMs. The prompt includes the focal method and test case with hint words "*The focal method:*" and "*The test case:*" to guide LLMs in generating test cases. The cosine learning rate is 1e−5. The batch size is eight. The maximum input sequence length is 1,024 tokens. If the number of input tokens exceeds the maximum number for the LLM, such as CodeBERT and AthenaTest, we truncate the input to meet the constraints. For StarCoder and CodeLlama7B, we adopt the Lora framework [29] to fine-tune LLMs, where lora_r is 8, lora_alpha is 16, and lora_dropout is 0.05. The type of task is set to "*CAUSAL_LM*". We set the seed for the random number generator to reduce the randomness in fine-tuning LLMs and make the generated results more stable and reliable. These settings are consistently applied to ensure a fair comparison between models fine-tuned with the original and filtered datasets. During the LLM inference phase, we use the greedy algorithm [36] to generate test cases, with a maximum output length of 512.

## 4.3   RQ1: The Distribution of Noise Data Types

In this RQ, we apply the noise data cleaning framework to detect noise and further analyze the quality of Methods2Test and Atlas datasets. Table 5 illustrates the distribution of each noise type in these datasets. We represent the percentage of noise identified by CleanTest. Since one focal method and test case pair may involve multiple types of noise, our tool repeatedly counts pairs with multiple noise types when calculating the frequency for each type. Since we count each type only once for the total frequency, the sum of individual type percentages exceeds the total noise percentage.

Overall, Methods2Test has the highest proportion of noise data (43.53%), followed by Atlas with 29.65%. We can observe that both datasets often contain multiple categories of noise. Methods2Test contains the most diverse noise categories. In Atlas, all noise categories are present except for the "no relevance" noise.

The Methods2Test dataset contains various types of noise affecting data quality, categorized into three main groups: syntactic noise, no relevance noise, and low coverage noise. The total noise observed across the entire dataset is significant, accounting for 43.53% of the total data, highlighting its substantial impact on data quality. Among the syntactic noise types, "Ambiguous Data Type" represents a relatively small portion, approximately 3.08%. "Unnecessary Annotations" is the most prevalent, making up 41.64% of the dataset, indicating a widespread issue with redundant data specifications. Other noise types account for smaller but noteworthy proportions, such as "Empty Exception Handling Statements" (0.68%), "Missing Implementation" (0.34%), and "Syntax Errors" (1.11%), highlighting varied challenges posed by these noise types. "Non-English Literal" noise accounts for 0.16%, the most minor proportion of the Methods2Test dataset, but it is still present. Furthermore, "No Relevance" and "Low Coverage" noises account for 12.70% and 3.95%, respectively. In Atlas, the percentages of "Ambiguous Data Type" (0.0051%), "Unnecessary Annotations" (0.0051%), "Missing Implementation" (0.03%), "Non-English Literal" (0.04%), and "No Relevance" (0%) noises are notably low. This low prevalence is due to the rigorous data preprocessing method. In particular, the "No Relevance" noise is 0%. Compared to the other noise types, "Syntax-Errors" noise accounts for a significant portion (28.74%). This figure is surprisingly high, especially since Atlas is sourced from high-quality GitHub repositories, many of which are well-maintained. A closer look suggests that this high percentage of noise could be due to the dataset's preprocessing. Partial or incomplete code fragments are included in the dataset.

Table 5. The Distribution of Noise in Methods2Test and Atlas Datasets

| Category of Noise Data | | Methods2Test | Atlas |
|---|---|---|---|
| Syntactic Noise | Ambiguous Data Type | 3.08% | 0.0051% |
| | Unnecessary Annotations | 41.64% | 0.0051% |
| | Empty Exception Handling Statement | 0.68% | 0.56% |
| | Missing-Implementation | 0.34% | 0.03% |
| | Syntax Errors | 1.11% | 28.74% |
| | Non-English Literal | 0.16% | 0.04% |
| No Relevance Noise | | 12.70% | 0% |
| Low Coverage Noise | | 3.95% | 0.77% |
| Total Noise | | 43.53% | 29.65% |

## 4.4  RQ2: The Effectiveness of the Noise Cleaning Framework

We investigate the impact of noise data test generation performance by fine-tuning four LLMs on two versions (original and filtered) of the Methods2Test and Atlas datasets. The "original" refers to the initial training dataset, while the "filtered" is the version cleaned by our noise-cleaning framework. It is important to note that the results presented here are based on Defects4J data, which is not included in the Methods2Test or Atlas datasets.

Table 6. The Test Generation Performance Comparisons of Four LLMs Fine-Tuned over Different Datasets

| Dataset | Model | Training Dataset | Training Time (Hours) | CodeBLEU | Syntactic Correct Rate | Compilation Passing Rate | Line Coverage | Branch Coverage |
|---|---|---|---|---|---|---|---|---|
| Methods2Test | CodeBERT | Original | 4.15 | 4.68% | 82.48% | 8.45% | 5.39% | 0.72% |
| | | Filtered | 2.75 (33.73% ↓) | 6.58% (40.60% ↑) | 81.44% (1.26% ↓) | 57.32% (578.34% ↑) | 6.78% (25.79% ↑) | 2.31% (220.83% ↑) |
| | AthenaTest | Original | 19.74 | 12.02% | 76.22% | 11.26% | 7.78% | 3.07% |
| | | Filtered | 13.08 (33.74% ↓) | 13.37% (11.23% ↑) | 83.20% (9.16% ↑) | 74.33% (560.12% ↑) | 7.84% (0.77% ↑) | 3.68% (19.87% ↑) |
| | StarCoder | Original | 120.33 | 3.53% | 78.93% | 41.67% | 11.84% | 5.79% |
| | | Filtered | 94.74 (21.27% ↓) | 17.06% (383.29% ↑) | 89.71% (13.66% ↑) | 37.25% (10.61% ↓) | 14.02% (18.41% ↑) | 5.83% (0.69% ↑) |
| | CodeLlama-7B | Original | 49.77 | 15.95% | 81.26% | 31.11% | 17.64% | 12.41% |
| | | Filtered | 43.27 (13.06% ↓) | 16.56% (3.82% ↑) | 91.98% (13.19% ↑) | 33.33% (7.14% ↑) | 22.76% (29.02% ↑) | 15.94% (28.44% ↑) |
| Atlas | CodeBERT | Original | 3.87 | 5.23% | 85.58% | 65.00% | 3.79% | 1.17% |
| | | Filtered | 3.50 (9.56% ↓) | 6.34% (21.22% ↑) | 96.15% (12.35% ↑) | 93.83% (44.35% ↑) | 5.58% (47.23% ↑) | 2.41% (105.98% ↑) |
| | AthenaTest | Original | 1.60 | 6.80% | 80.94% | 77.78% | 4.65% | 2.13% |
| | | Filtered | 1.50 (6.25% ↓) | 6.98% (2.65% ↑) | 90.81% (12.19% ↑) | 80.63% (3.66% ↑) | 5.01% (7.74% ↑) | 2.51% (17.84% ↑) |
| | StarCoder | Original | 27.20 | 11.53% | 42.19% | 60.11% | 3.46% | 2.75% |
| | | Filtered | 22.28 (18.09% ↓) | 12.87% (11.62% ↑) | 61.73% (46.31% ↑) | 69.23% (15.17% ↑) | 3.95% (14.16% ↑) | 2.98% (8.36% ↑) |
| | CodeLlama-7B | Original | 13.64 | 9.06% | 60.63% | 75.00% | 5.27% | 2.78% |
| | | Filtered | 12.81 (6.09% ↓) | 9.15% (0.99% ↑) | 63.48% (4.70% ↑) | 85.71% (14.28% ↑) | 6.33% (20.11% ↑) | 3.47% (24.82% ↑) |

Table 6 shows the performance of four LLMs, with the best results highlighted in bold. Overall, filtering noise data from the Methods2Test and Atlas datasets positively impacts test generation performance across the four LLMs. Fine-tuning the four LLMs with the filtered datasets results in improvements in CodeBLEU (by 30.91%, 6.94%, 197.45%, and 2.41%), syntactic correctness rate (by 5.55%, 10.68%, 29.99%, and 8.95%), compilation passing rate (by 311.35%, 281.89%, 2.28%, and 10.71%), line coverage (by 36.51%, 4.26%, 16.29%, and 24.57%), and branch coverage (by 163.41%, 18.86%, 4.53%, and 26.63%), respectively. The fine-tuning time is saved by 21.65%, 19.99%, 19.68%, and 9.57%. Besides, we also observe the following findings:

(1) The impact of noises on the Methods2Test dataset is greater than the Atlas dataset. Fine-tuning the four LLMs with the filtered Methods2Test dataset results in average improvements of 109.74% in CodeBLEU, 8.69 in syntactic correctness rate, 283.75 in compilation passing rate, 18.50 in line coverage, and 67.46 in branch coverage. The fine-tuning time is reduced by 25.45% on average. For the Atlas dataset, the five metrics improve by 9.12%, 18.89%, 19.37%, 22.31%, and 39.25%, on average. The fine-tuning time is saved by 10.00%.

(2) The relative performance improvements of the four LLMs vary. For the five metrics, StarCoder achieves the highest improvements in CodeBLEU (197.45%) and syntactic correctness rate (29.99%). CodeBERT shows the greatest gains in compilation passing rate (311.35%), line coverage (36.51%), and branch coverage (163.41%). These differences in performance are likely due to variations in the LLMs' pre-training methods, architectures, and training data, leading to different levels of improvement after fine-tuning on the same dataset.

(3) When using the filtered Methods2Test dataset, the syntactic correctness rate of CodeBERT decreases by 1.26%. The compilation passing rate of StarCoder also worsens by 10.61%. This may be because the filtered dataset includes only high-quality examples, which can present more challenging coding scenarios. As a result, achieving syntactic correctness and successful compilation becomes more difficult, as these tasks require precise code syntax and dependency handling. However, high-quality datasets also enable models to better grasp complex code logic and patterns, improving broader coverage and semantic understanding.

Fig. 10. An Example of Generated Test Cases Using LLMs Fine-Tuned with Original and Filtered Datasets

To qualitatively illustrate the impact of noise on test generation models, we present a case generated by CodeLlama7B fine-tuned on both the original and filtered Methods2Test datasets. As shown in Fig. 10, the test case generated by CodeLlama7B fine-tuned on the filtered dataset (short for CodeLlama7B-filtered) tends to be broader and more accurate than the test case generated by CodeLlama7B fine-tuned on the original dataset (short for CodeLlama7B-origin).

The focal method "`between()`" is a time processing function that calculates the time difference between two points. The test case generated by CodeLlama7B-origin verifies whether the method correctly throws an "`IllegalArgumentException`" when given a null input. In contrast, the test cases generated by CodeLlama7B-filtered not only handle an extreme case (the time difference from 2004 to 1970) but also test the method's ability to handle different time units, such as milliseconds and minutes. Further, the filtered test case includes multiple assertion statements, verifying the correctness of the time calculation in different scenarios, such as using different time units or modifying the end time (e.g., adding one minute). This case also tests cases where the start and end times are identical, an essential check for verifying the accuracy of the time calculation logic. By eliminating irrelevant or incorrect data during the filtering process, the model is able to generate more comprehensive and accurate test cases, leading to broader and more detailed coverage in the CodeLlama7B-filtered results. Besides, it also verifies the correctness of the function under normal and boundary conditions, demonstrating the reliability and robustness of the code in multiple scenarios by using different time units. Therefore, CodeLlama7B-filtered's test cases are superior to CodeLlama7B-origin's due to their comprehensiveness and testing depth.

## 4.5 RQ3: The Impact of Each Filter Component

In CleanTest, we design three filters to improve the quality of the Methods2Test and Atlas datasets. We evaluate the effectiveness of each filter component through ablation experiments, disabling one of the three filter components in turn. After fine-tuning with the derived filtered Methods2Test dataset, we observe the model performance and compare it with previous results. If the performance declines compared to when all three filters are enabled, we can infer a positive impact of the disabled component on the framework's effectiveness.

Table 7. The Performance Comparisons of Different Filter Component in the Methods2Test Dataset

| Training Dataset | CodeBLEU | Syntactic Correctness Rate | Compilation Passing Rate | Line Coverage | Branch Coverage |
|---|---|---|---|---|---|
| **All Filter** | **16.56%** | **91.98%** | 33.33% | **22.76%** | **15.94%** |
| **Syntactic Filter & Relevance Filter** | 16.08% | 89.53% | **36.96%** | 22.70% | 15.30% |
| **Only Syntactic Filter** | 15.68% | 91.03% | 29.55% | 19.10% | 13.39% |
| **No Filter (i.e., Origin)** | 15.95% | 81.26% | 31.11% | 17.64% | 12.41% |

Table 7 presents the performance of CodeLlama7B fine-tuned on different filtered Methods2Test datasets. We observe that removing any filter results in lower performance scores. Specifically, when the model-based coverage filter is removed, CodeBLEU, syntactic correctness rate, line coverage, and branch coverage decrease by 2.99%, 2.74%, 0.26%, and 4.18%, respectively. However, the compilation passing rate increases by 9.82%. The reason is that without the coverage filter, the model tends to generate simpler or less comprehensive code, which is easier to compile but lacks the necessary complexity for thorough testing. As a result, while more code passes compilation, it does not perform as well in terms of correctness and coverage. After removing the relevance filter, the performance of CodeLlama7B drops by 2.55% in CodeBLEU, 25.08% in compilation passing rate, 18.85% in line coverage, and 14.26% in branch coverage. Besides, the observed decrease in CodeBLEU and compilation passing rate for the "Only Syntactic Filter" variant, despite the slight improvement in other metrics, could be attributed to the following reasons: (1) While the syntactic filter ensures that the generated code adheres to proper syntax, it does not necessarily guarantee that the code is semantically correct or relevant to the intended task. Without additional relevance filtering, LLMs might produce syntactically correct code that fails to compile correctly or is not meaningful, leading to a lower compilation passing rate. (2) The focus on syntactic accuracy could have reduced the diversity or flexibility of the generated code, limiting its ability to meet broader criteria evaluated by CodeBLEU or pass compilation checks, which often require not just correct syntax but also correct logic and compatibility with the programming environment.

## 5 Discussion

In this section, we first discuss the effectiveness of CleanTest in bug detection. Then, we introduce the threats to validity.

### 5.1 The Performance on Bug Detection

*5.1.1 The Bug Detection Performance of Different LLMs on Different Datasets.* We follow previous research [16] and also adopt the Defects4J benchmark [21, 34] to evaluate the effectiveness of the generated test code in bug detection. Specifically, we first use the fine-tuned LLMs to generate test cases. Then, we replace the manually written test cases with the test code generated by fine-tuned LLMs. Finally, we run the replaced test cases on the buggy and fixed version and record the number of test cases that fail on the buggy version but pass on the fixed version. This ensures reliable bug identification and robust evaluation of generated test cases. To evaluate the performance of the four LLMs in bug detection, we adopt the number of detected bugs (short for #detected bugs) as the metric.

Table 8 shows the bug detection performance of different LLMs after fine-tuning with the original and filtered Methods2Test datasets, with the best results in bold. We can see that removing noise in the Methods2Test dataset increases the number of detected bugs by 24.23% on average. All four models show improvements in bug detection performance, with AthenaTest showing the largest improvement. We believe the reason is that a high-quality dataset provides the model with a clearer relationship between code behavior and expected results, allowing it to explore different logical conditions within the focal method.

Table 8. The Bug Detection Performance Comparisons of Four LLMs Fine-Tuned over Different Datasets

| Model | CodeBERT | | AthenaTest | | StarCoder | | CodeLlama7B | |
|---|---|---|---|---|---|---|---|---|
| Training Dataset | Original | Filtered | Original | Filtered | Original | Filtered | Original | Filtered |
| #Detected Bugs | 14 | **16** | 17 | **24** | 15 | **18** | 14 | **17** |

To further understand the effect of CleanTest in the bug detection task, we manually review the test cases generated by CodeLlama7B-filtered and CodeLlama7B-origin. By comparing the test cases generated by the two, we find that the test code generated by CodeLlama7B-filtered contains more comprehensive and accurate assertion statements. Fig. 11 presents an example, where the focal method "tanh()" is a hyperbolic tangent function collected from the Math project [47]. Line 2 has a potential bug because it does not consider the case where the real part is infinite. The "tanh()" function should check the real and imaginary parts separately. The assertion statement in line 7 of the test case generated by CodeLlama7B-filtered can detect this potential bug in the focal method. This statement assumes a complex number object whose real part is infinity and whose imaginary part is 1. "nanZero" is a complex number object with a "NaN" real part and a 0 imaginary part. The assertion statement fails to ensure that the "tanh()" function returns the expected result, thereby discovering the bug in the focal method. Besides, this test case also includes other types of inputs. For example, line 3 verifies that the real part is close to 0.000001 within an accuracy of 1.0e−5. Line 5 aims to check if the imaginary part is negative infinity. The "tanh()" function should return "Complex.NaN".

In contrast, the test case generated by CodeLlama7B-origin only tests a common complex input (i.e., "(0.25, 0.5)"), which does not cover the above boundary conditions and extreme cases. The bug in the focal method "tanh()" cannot be detected. The test case generated by CodeLlama7B-filtered is more comprehensive and robust than that generated by CodeLlama7B-origin. Through the tests CodeLlama7B-filtered generated, the correctness and robustness of the focal method can be more effectively verified.

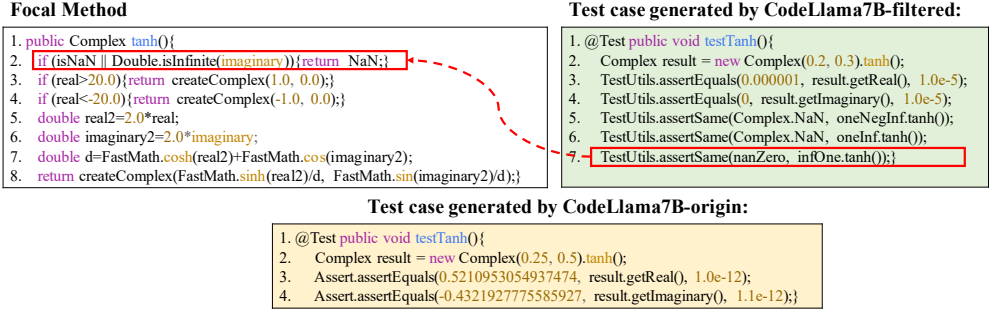

Fig. 11. An Example of Generated Test Cases Using CodeLlama7B Fine-Tuned with Original and Filtered Datasets

Table 9. The Bug Detection Performance Comparisons of Different Filter Component

| Training Dataset | No Filter (i.e., Origin) | Only Syntactic Filter | Syntactic Filter & Relevance Filter | All Filtered |
|---|---|---|---|---|
| #Detected Bugs | 14 | 14 | 15 | **17** |

*5.1.2 The Impact of Each Filter Component on Bug Detection Performance.* To further explore the impact of different filtering components in bug detection, we conduct an ablation experiment by observing the performance of CodeLlama7B after fine-tuning the Methods2Test datasets filtered with different filters and compare it with the performance of Codellama7B-origin. The results of our comparisons are presented in Table 9. When CleanTest only includes the syntactic filter, the number of detected bugs remains unchanged. After adding the relevance filter, the number of detected bugs increases by one. With all three filters, the number of detected bugs increases to seventeen. This indicates that high-quality datasets, especially those with high branch coverage, help models generate tests to reveal bugs. By learning from such high-quality datasets, models become adept at generating tests that cover more conditional paths, thereby uncovering bugs that might otherwise remain hidden.

## 5.2 Threats to Validity

**(1) Syntactic Filter.** Our heuristic rules only cover Java datasets, posing a threat to the generalizability of our framework. Although we believe all test generation datasets can benefit from our framework, we will focus on datasets for other programming languages in future work. **(2) Manual Annotation.** The threat from manual annotation in the open card sorting method is acknowledged. To reduce this threat, we establish a labeling team and perform multiple rounds of labeling to ensure that all participants achieve conceptual coherence. The method involves multiple assessors, helps mitigate individual biases, and ensures that the final taxonomy is a balanced representation of different viewpoints. **(3) Random Sampling.** We adopt random sampling in constructing the taxonomy of noise data, which may lead to incomplete results. We plan to enlarge the analyzed dataset and inspect whether new types of noise emerge. **(4) Coverage Computation.** In the experimental section, we use Cobertura [9] (rather than JaCoCo [32]) to calculate coverage. This choice is fundamentally rooted in our experimental design, specifically our reliance on Defects4J as the evaluation benchmark. Defects4J, widely used for evaluating automated test generation models, utilizes Cobertura to calculate coverage. By aligning our evaluation tools with those used by Defects4J, we ensure consistency in coverage metrics, facilitating more accurate and direct comparisons of our results with existing studies using Defects4J [61, 76]. **(5) Validation Dataset.** In this work, we use Defects4J as the validation dataset to evaluate the impact of noise data on four LLMs. These LLMs are trained on a large corpus of code collected from GitHub [18, 38, 51, 61]. Since Defects4J contains public GitHub repositories, it is unclear if these repositories are part of the Methods2Test and Atlas datasets, raising the risk of overfitting and biased evaluation results. To mitigate this risk, we filter the training data by removing any code that shares the same function name with the code in Defects4J. Instead, we focus on fine-tuning the models with filtered and unfiltered data, and the relative performance improvement provides valuable insights. **(6) Code Coverage Dataset.** In this work, the coverage filter is fine-tuned using a dataset provided in a previous study from GitHub. We use the coverage filter to predict the branch coverage of the Methods2Test dataset, also from GitHub. Hence, there is a similar threat to that posed by the validation dataset. To avoid data leakage, we delete data whose focal method names in this dataset match the code names in the Methods2Test dataset.

## 6 Related Work

### 6.1 Unit Test Generation.

Various techniques have been developed to generate unit tests. In the literature, traditional test generation models can be categorized into three types based on the techniques used: random-based approaches [43], constraint-based approaches [12, 17, 72], and search-based approaches [7, 14, 25]. Despite achieving promising results, these approaches struggle with the path explosion problem in symbolic execution or might not effectively explore the vast input space to detect intricate bugs [3]. Besides, these approaches also struggle in terms of explainability and readability [65]. Recently, LLMs have exhibited remarkable performance across a wide range of coding-related tasks [20, 22, 30, 70, 71, 73]. Some studies also utilize LLMs to generate test cases [1, 37, 61]. For instance, Alagarsamy et al. [1] pre-trained and fine-tuned the LLM with the focal method and assertion statements to generate test cases. Tufano et al. [61] utilized a large Java corpus to pre-train the LLM and fine-tuned it with a translation task to generate unit tests. Hashtroudi et al. [26] fine-tuned the LLM with existing developer-written tests to generate domain adaptation tests. Rao et al. [49] considered the mapping between code and test files as the pre-training signal to train a GPT-style LLM and generate tests. Steenhoek et al. [56] utilized multiple static quality metrics as rewards and adopted the reinforcement learning strategy to optimize the LLM and generate tests. Besides, Schafer et al. [54] conducted an empirical evaluation of using LLMs for unit test generation. Guilherme et al. [24] investigated the capability of ChatGPT to generate unit tests.

Our work differs from prior research in several key aspects. First, we study multiple LLMs fine-tuned with a high-quality test generation dataset. Second, we propose a noise-cleaning framework to filter out noise data and improve the quality of the test generation dataset.

### 6.2 Data Quality in Software Engineering.

Since software engineering data and artifacts are usually collected via mining software repositories [8], they are not explicitly generated for research, and their quality can vary. To address this, many researchers have focused on improving data quality to enhance the accuracy and robustness of code models [39]. For example, Croft et al. [11] discussed four data quality attributes (i.e., accuracy, completeness, consistency, and timeliness) for software vulnerability datasets. Sun et al. [58] proposed a data-cleaning framework for neural code search to improve the quality of code search datasets. Similarly, Shi et al. [55] conducted systematic research to assess and improve the quality of four code summarization datasets. Wu et al. [68] conducted case studies on five publicly available security bug report prediction datasets and improved their labeling accuracy by manually analyzing each bug report.

While these studies focus on cleaning and improving datasets, there is a lack of in-depth analysis regarding the quality of unit test generation datasets. Our study bridges this gap with a large-scale analysis of data preprocessing errors and low-quality focal methods and test cases. We design a noise-cleaning framework with three filters to improve dataset quality and investigate the performance variation of existing models on the filtered dataset.

## 7 Conclusion and Future Work

In this paper, we construct a taxonomy of noise data for test generation, identifying eight types of noise. Our study demonstrates the prevalence of noise in the Methods2Test dataset. We propose a noise-cleaning framework to detect and filter out noise. The framework includes three filters: the syntactic filter, the relevance filter, and the coverage filter. The syntactic filter uses heuristic rules to filter out focal methods and test cases with syntactic anomalies. The relevance filter selects datasets where the function signatures, parameter types, and the number of parameters between the focal methods and the called functions in test cases match. The model-based coverage filter refines the dataset to retain high branch coverage. It is based on the fine-tuned CodeGPT over a pre-collected coverage dataset and uses this information to select test cases with higher code coverage. Experiments show that our noise-cleaning framework can improve the performance at least 39% in branch coverage and 21% in the number of detected bugs. The fine-tuning time is saved by at least 10%. In the future, we plan to extend our study to detect noise in other types of datasets (e.g., code generation and code review).

## 8 Data Availability

The replication of this paper is publicly available [44].

### Acknowledgments

This research is supported by the National Key R&D Program of China (No. 2024YFB4506400), CCF-Huawei Populus Grove Fund, and the National Research Foundation, under its Investigatorship Grant (NRF-NRFI08-2022-0002). Any opinions, findings and conclusions or recommendations expressed in this material are those of the author(s) and do not reflect the views of National Research Foundation, Singapore.